\documentclass[12pt]{article}
\usepackage[dvips]{graphicx}
\usepackage{mathbbol}
\usepackage{amsmath}
\usepackage{amssymb,amsfonts}

\newcommand{\eq}{\begin{equation}}
\newcommand{\en}{\end{equation}}
\newcommand{\eqa}{\begin{eqnarray}}
\newcommand{\ena}{\end{eqnarray}}

\begin{document}

\begin{titlepage}
\vskip0.5cm
\begin{flushright}
\end{flushright}
\vskip0.5cm
\begin{center}
{\Large\bf The three-dimensional XY universality class: \\
A high precision Monte Carlo estimate of the universal amplitude 
ratio $A_+/A_-$}
\vskip 0.3cm
{\Large\bf }
\vskip 0.3cm
\end{center}
\vskip 1.3cm
\centerline{
Martin Hasenbusch
}
\vskip 0.4cm
\centerline{\sl Dipartimento di Fisica dell'Universit\`a di Pisa
 and I.N.F.N.,} 
\centerline{\sl I-56127 Pisa, Italy}
\vskip 0.3cm

\vskip 0.4cm
\begin{abstract}
We simulate the improved three-dimensional two-component $\phi^4$ model on 
the simple cubic lattice 
in the low and the high temperature phase for reduced temperatures
down to $|T-T_c|/T_c \approx 0.0017$ on lattices of a size up to $350^3$.
Our new results for the internal energy and the specific heat are  
combined with the accurate estimates of $\beta_c$ and data for the 
internal energy and the specific heat at $\beta_c$ recently obtained in 
ref.~\cite{we2006}. We find
$R_{\alpha} = (1-A_+/A_-)/\alpha = 4.01(5)$, where $\alpha$ is the critical
exponent of the specific heat and $A_{\pm}$ is the amplitude
of the specific heat in the high and the low temperature phase, respectively.
\end{abstract}
\end{titlepage}

\section{Introduction}
The renormalization group (RG) theory of critical phenomena
classifies continuous phase transitions into so called universality classes, 
which are characterized by the dimension of the system, 
the range of the interaction and the symmetry of the order parameter;
see e.g. \cite{WiKo,Parisi,cardy,Fisher98}.
At continuous phase transitions, thermodynamic quantities follow power laws.
E.g. the specific heat behaves as
\begin{equation}
\label{specific_heat}
 C \simeq A_{\pm}  |t|^{-\alpha} (1 + c_{\pm} |t|^{\theta}+ ...) + c_{ns} \;, 
\end{equation}
where $\alpha$ is the critical exponent of the specific heat, $\theta$ the
exponent of the leading correction to scaling, 
$A_{+},A_{-}$,$c_{+}$,$c_{-}$ are amplitudes in the high and low 
temperature phase, respectively.
The reduced temperature $t=(T-T_c)/T_c$ gives the distance from
the critical temperature $T_c$. 
$c_{ns}$ is the analytic background, which has
to be taken into account when $\alpha \le  0$, as it is the case here.
Following the RG theory, critical exponents are universal, which means
that they take exactly the same value for any system within a given
universality class. Most recent estimates are $\alpha=-0.0151(3)$ and 
$\theta=0.527(13)$ \cite{we2006}.  For reviews on theoretical and 
experimental results see \cite{evap,ourXY}.
In addition to the critical exponents,
amplitude ratios like $A_{+}/A_{-}$ and $c_{+}/c_{-}$ are universal, 
while the individual values of
$A_{+}$,$A_{-}$,$c_{+}$ and $c_{-}$ depend on the microscopic details of 
the model.

The three-dimensional XY universality class is of particular interest, 
since the
$\lambda$-transition of $^4$He is supposed to share this universality class.
The experimental study of this transition provides by far the most precise
experimental results for universal quantities like critical exponents and 
amplitude ratios.
Thus this transition gives us an unique opportunity to test the
ideas of the renormalization group and to benchmark  theoretical
methods. Most recent experiments with $^4$He were carried out during a spacelab
mission \cite{Lipa-etal-00,lipa2003}. The condition of micro-gravity avoids
the broadening of the transition due to the gravitational field and hence
allows to access reduced temperatures down to $t \approx 5 \times 10^{-10}$.
The most recent analysis \cite{lipa2003} of the spacelab 
data gives $\alpha=-0.0127(3)$
and $A_{+}/A_{-}=1.053(2)$ or 
\begin{equation}
\label{rspacelab}
R_{\alpha}=\frac{1-A_{+}/A_{-}}{\alpha}=4.154(22) \;\;. 
\end{equation}
Note that $R_{\alpha}$ is much less correlated with the value of $\alpha$ 
than $A_{+}/A_{-}$ itself. This result can be compared with other experimental
results $R_{\alpha}=4.194(19)$ \cite{ahlers}, 
$R_{\alpha}=4.57(40)$ \cite{lich1}, $R_{\alpha}=3.98(2)$ \cite{lich2},
field theoretic estimates $R_{\alpha}=4.433(77)$ \cite{SD-03},
Monte Carlo simulations $R_{\alpha}=4.20(5)$ \cite{CEHMS-02} and 
high temperature expansions combined with the equation of state
$R_{\alpha}=4.3(2)$ \cite{we2006}. Here we make no attempt to give a complete 
overview of theoretical results, we just try to give the most recent and 
hopefully most accurate result for each of the methods.

Notice that some of these estimates of $R_{\alpha}$ are not consistent 
among each other.
This could be interpreted as a violation of universality; however we regard
it as more likely that systematic errors are underestimated by some of the 
authors. Here we make an effort to keep errors, in particular the 
systematic error due to corrections to scaling, under control. 

We studied the $\phi^4$ model on a cubic lattice
with periodic boundary conditions in each of the directions. The lattice size
is $L^3$, where $L$ is the linear extension of the lattice. The lattice 
spacing is set to $a=1$.
The classical Hamiltonian is given by
\begin{equation}
 H = - \beta \sum_{<x,y>} \vec{\phi}_x \cdot \vec{\phi}_y
   + \sum_{x} \left[\vec{\phi}_x^2 + \lambda (\vec{\phi}_x^2 -1)^2   \right]
\;\;,
\end{equation}
where the field variable $\vec{\phi}_x$ is a vector with two real
components. $<x,y>$ denotes a pair of nearest neighbour sites.
The partition function is given by
\begin{equation}
 Z(\beta,\lambda) = \int \mbox{D} [\phi] \; \exp(- H(\beta,\lambda,{\phi})) \;,
\end{equation}
where $\int \mbox{D} [\phi]$ denotes the $2 L^3$ dimensional integral over the
field variables.
In the limit $\lambda \rightarrow \infty$ the classical XY (plane rotator)
model is recovered. Corrections to scaling amplitudes such as $c_{\pm}$  
of eq.~(\ref{specific_heat}) are  functions of the parameter $\lambda$. 
It has been demonstrated that there exists a value $\lambda^*$ at which 
the amplitudes of the leading correction to scaling vanish.
The most recent numerical
estimate is $\lambda^* = 2.15(5)$ obtained in ref. \cite{we2006}. 
Previous estimates are $\lambda^* = 2.07(5)$
and $\lambda^* = 2.10(6)$ given in refs. \cite{ourXY,tiborandI}, respectively.

Here we shall analyse data for the specific heat and the energy density at 
$\lambda = 2.1$ and $\lambda = 2.2$. 
In ref. \cite{we2006} accurate estimates of the inverse critical temperature
at various values of $\lambda$ are given. In the following we shall use
\begin{eqnarray}
\label{betac}
\beta_c&=&0.5091503(3)[3] \;\;\;\;\; \mbox{at} \;\;\; \lambda=2.1 \;\; , 
\nonumber \\
\beta_c&=&0.5083355(3)[4] \;\;\;\;\; \mbox{at} \;\;\; \lambda=2.2 \;\; .
\end{eqnarray}
The number in the parentheses gives the statistical error, while
the number in brackets is an estimate of possible systematic errors.


The outline of the paper is the following.  In the next section we
define the energy and the specific heat for our lattice model. We 
summarize the predictions of the RG-theory for the free energy density
and the specific heat. 
In section \ref{numcrit} we present our numerical results. 
First we analyse the finite size behaviour of the 
energy density and the specific heat at the critical temperature.
Then we discuss our results for the low and the high temperature phase.
Finally an estimate for $R_{\alpha}$ is obtained by fitting these data 
to the expected power law behaviour.

\section{Critical behaviour of the energy and the specific heat}
The so called reduced free energy density  is defined by
\begin{equation}
f(\beta,\lambda) = - \frac{1}{V} \ln Z(\beta,\lambda) \;\;, 
\end{equation}
which is more convenient for our purposes than the usual 
$-\frac{T}{V} \; \ln Z$.

We  define the energy density as
\begin{equation}
 E= - \frac{\partial f}{\partial \beta} = 
 \frac{1}{V} \sum_{<xy>} \langle \phi_x \phi_y \rangle
\end{equation}
and the specific heat as
\begin{equation}
 C= - \frac{\partial^2 f}{\partial \beta^2}=
\frac{1}{V} \left( \langle (\sum_{<xy>} \phi_x \phi_y)^2 \rangle - 
		       \langle \sum_{<xy>} \phi_x \phi_y \rangle^2 \right)
\;\;.
\end{equation}
These definitions differ by factors $-1$ and $\beta^2$
from standard textbook definitions. 
these factors cancel in the amplitude ratio.
Now, let us summarize the predictions of the RG-theory for the free energy
density. First, the free energy is split into a singular and non-singular 
part
\begin{equation}
f = f_{ns} + f_s \;\; .
\end{equation}
Let us first discuss the finite size scaling (FSS) behaviour.
The singular part of the free energy density depends on the parameters of
the model and the lattice size $L$. Renormalization group
(see e.g.~\cite{Privman-90,SS-00}) 
predicts  for a system with periodic boundary conditions
\begin{eqnarray}
f_s(u_t, u_h, \{u_i\},L)  =
 L^{-d}
\Phi( L^{y_t} u_t, L^{y_h} u_h, \{L^{y_i} u_i\}) \;\;,
\label{FscalL}
\end{eqnarray}
where $u_t\equiv u_1$, $u_h\equiv u_2$, $\{u_i\}$ with $i\geq 3$ are
the scaling fields (which are analytic functions of the Hamiltonian
parameters) associated with the reduced temperature $t$
($u_t\sim t$), the magnetic field $h$ ($u_h\sim h$), and the other
irrelevant perturbations with $y_i<0$, respectively. 
$d$ is the dimension of the system; in our case $d=3$.
The exponent of the thermal scaling field is 
related with the critical exponent of the correlation length: $y_t=1/\nu$.
Furthermore the hyperscaling relation $\alpha=2-d \nu=2-d/y_t$ holds.

Let us consider the scaling of the energy density and the specific heat 
at the critical point $u_t=0$ and $u_h=0$.
Taking the derivative with respect to $\beta$ and Taylor-expanding 
in $u_i$ with $i\geq 3$  we get 
\begin{equation}
\label{fssE}
 E_s(L,\beta_c) =e_s L^{-d+y_t}  (1+d_1 L^{y_3} + d_2 L^{y_4} +...)
\end{equation}
and 
\begin{equation}
\label{fssC}
 C_s(L,\beta_c) =c_s L^{-d+2 y_t}  (1+f_1 L^{y_3} + f_2 L^{y_4} +...)
\end{equation}
for the singular parts of the energy density and the specific heat, 
respectively.
The numerical values of the correction exponents are 
$-y_3=\omega=0.785(20)$ \cite{we2006} and $-y_4=\omega_2\approx 1.8$
\cite{NR-84,we2006}.

In the thermodynamic limit, for vanishing external field $h=0$, the 
singular part of the free energy density behaves as
\begin{equation}
\label{fsthermodynamic}
f_s = \tilde a_{\pm} |t|^{2-\alpha} \; 
\left(1+ c_{1,1,\pm} |t|^{\Delta} + c_{1,2,\pm}  |t|^{2 \Delta}  
               + d_1 t + c_{2,1,\pm} |t|^{ \Delta_2} + ...\right)
\end{equation}
with $\Delta =\nu \omega$ and $\Delta_2 =\nu \omega_2$.  
In the following we define the  reduced temperature by $t=\beta-\beta_c$.
Predictions for the singular parts of the 
energy density and the specific heat can be easily
derived from eq.~(\ref{fsthermodynamic}) by taking the first and second 
derivative with respect to $\beta$. One gets
$
e_s = a_{\pm}  |t|^{1-\alpha} \; \left(1+ ... \right)
$
and 
$
c_s = A_{\pm}  |t|^{-\alpha} \; \left(1+ ... \right)
$
with
\begin{equation}
\frac{A_{+}}{A_{-}} = - \frac{a_{+}}{a_{-}} = \frac{\tilde a_{+}}{ \tilde a_{-}}
\;\;.
\end{equation}

\section{The Numerical results}
\label{numcrit}
Let us first sketch the strategy of our numerical analysis.  
We parametrise the non-singular part of the free energy by its
Taylor-expansion in the reduced temperature $t=\beta-\beta_c$:
\begin{equation}
\label{taylorfns}
 f_{ns}(t) = f_{ns}(0) + e_{ns} t + \frac{1}{2} c_{ns} t^2 + ... \;\;.
\end{equation}
In a first step of the analysis we determine $e_{ns}$ and $c_{ns}$ from 
the finite size behaviour of the energy density and the specific heat 
at the critical temperature. Then, in the analysis of the data for 
the thermodynamic limit at $\beta\ne \beta_c$, these estimates 
for $e_{ns}$ and $c_{ns}$ are used as input. In this respect, we essentially 
follow ref.~\cite{CEHMS-02}.

Note that the amplitudes of the
leading correction to scaling are small at $\lambda=2.1$ and 
$\lambda=2.2$. Therefore we shall not take into account the leading  
correction to scaling in our fits. 
The systematic error introduced this way can be estimated by comparing the 
final results obtained for $\lambda=2.1$ and $\lambda=2.2$.

\subsection{The energy density and the specific heat at the critical
temperature}
We analyse data for the energy density and the specific heat obtained on 
lattices of a linear size up to $L=128$. 
The simulations are performed for some $\beta_s \approx \beta_c$. We have 
computed the Taylor coefficients of the energy density at $\beta_s$
up to $2^{nd}$ order. This allows us to compute the energy density and the 
specific heat in a sufficiently large neighbourhood of $\beta_s$.  

Most of the data for $\lambda=2.1$ were generated already for ref. \cite{ourXY}.
Some additional data, in particular data for $L=128$, were generated 
more recently for ref. \cite{we2006}. In the case of $\lambda=2.2$  the data
for $L \le 16$ and for $L=128$ were generated  for ref. \cite{we2006}.
Here we have added new data for $L=24,32,48$ and $64$.  In the Monte Carlo
simulation a mixture of local and wall-cluster \cite{Pinnetal}
updates was used. For details we refer the reader to refs. \cite{ourXY,we2006}.

First we have analysed the energy density at the central estimates of $\beta_c$
given in eq.~(\ref{betac}). In addition, to propagate the error of $\beta_c$, 
we have analysed the energy density at a slightly shifted value for $\beta_c$.
We fitted our data using the ansatz
\begin{equation}
\label{simpleE}
 E = e_{ns} + e_s L^{-d+1/\nu} \;
\end{equation}
without any correction term.  We have performed these fits fixing $\nu=0.6717$,
which is the central estimate of ref. \cite{we2006}, corresponding to 
$\alpha=-0.0151$. 
In addition we performed fits with slightly different 
values of $\nu$ to determine the dependence of the result for $e_{ns}$
on the input value of $\nu$. 
For $\nu=0.6717$, these fits lead to a $\chi^2$/d.o.f. $\approx 1$ 
starting from $L_{min}=6$, where $L_{min}$ is the minimal lattice size 
that has been included into the fit. Furthermore, the result
for $e_{ns}$ is very stable when $L_{min}$ is increased. Also the statistical
error of $e_{ns}$ increases only slowly with increasing $L_{min}$. 
Hence we regard the estimate of $e_{ns}$ obtained from these a fits as reliable.
Being very conservative, we take our final result from fits with $L_{min}= 24$.
Our results can be summarized as 
\begin{equation}
\label{e0l2.1}
 e_{ns} = 0.913213(5) + 20 \times (\beta- 0.5091503)
                    + 5 \times 10^{-7} \times (1/\alpha+1/0.0151)  
\end{equation}
for $\lambda=2.1$ and
\begin{equation}
\label{e0l2.2}
 e_{ns} = 0.913585(5) + 20 \times (\beta-0.5083355)
                     + 6 \times 10^{-7} \times (1/\alpha+1/0.0151) 
\end{equation}
for $\lambda=2.2$. For the later use it is convenient to parametrise
the dependence of the result on the input parameter $\nu$ in terms 
of $1/\alpha$. 

\begin{figure}[tp]
\begin{center}
\scalebox{0.5}
{
\includegraphics{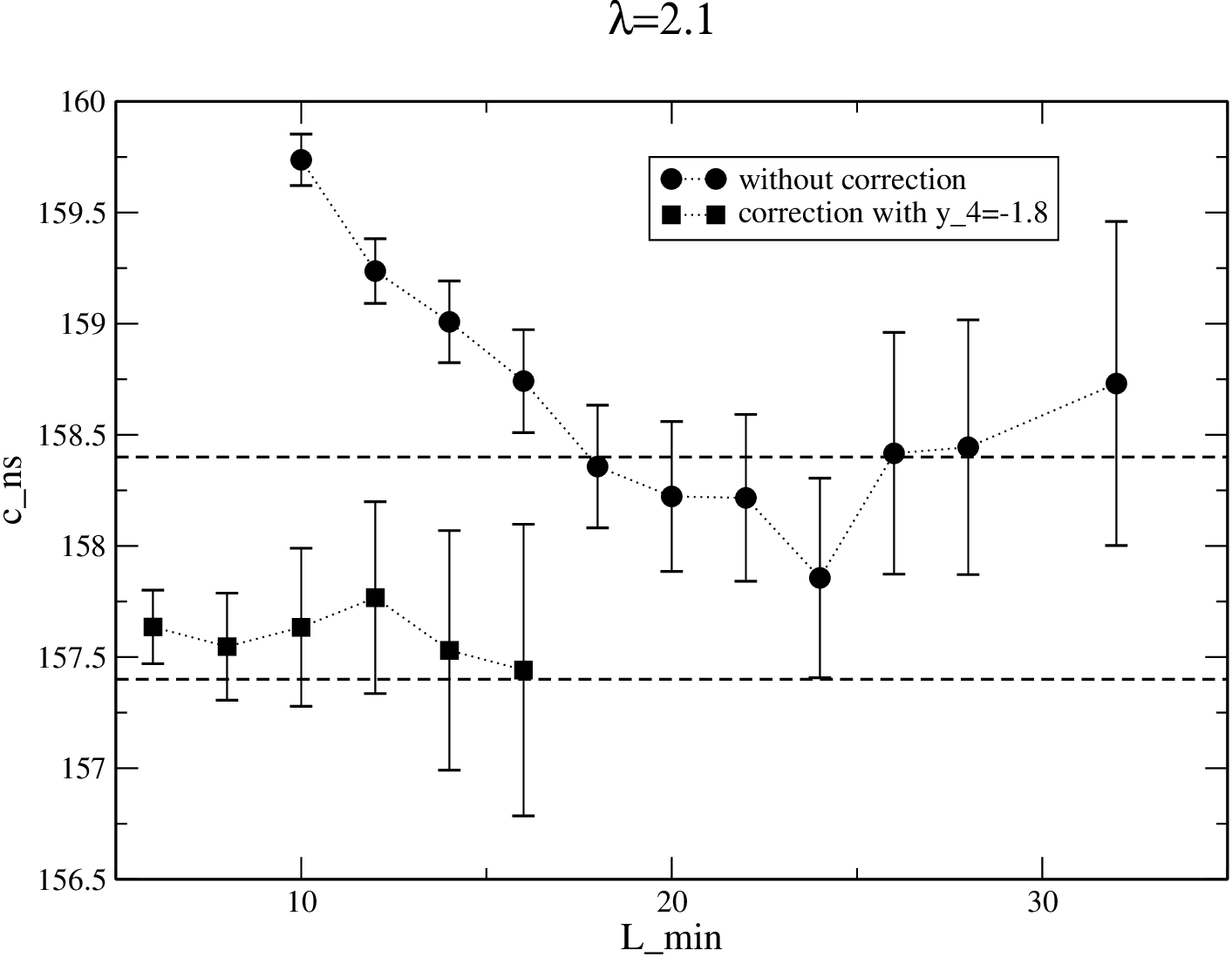}
} \\ 
\vskip0.6cm
\scalebox{0.5}
{
\includegraphics{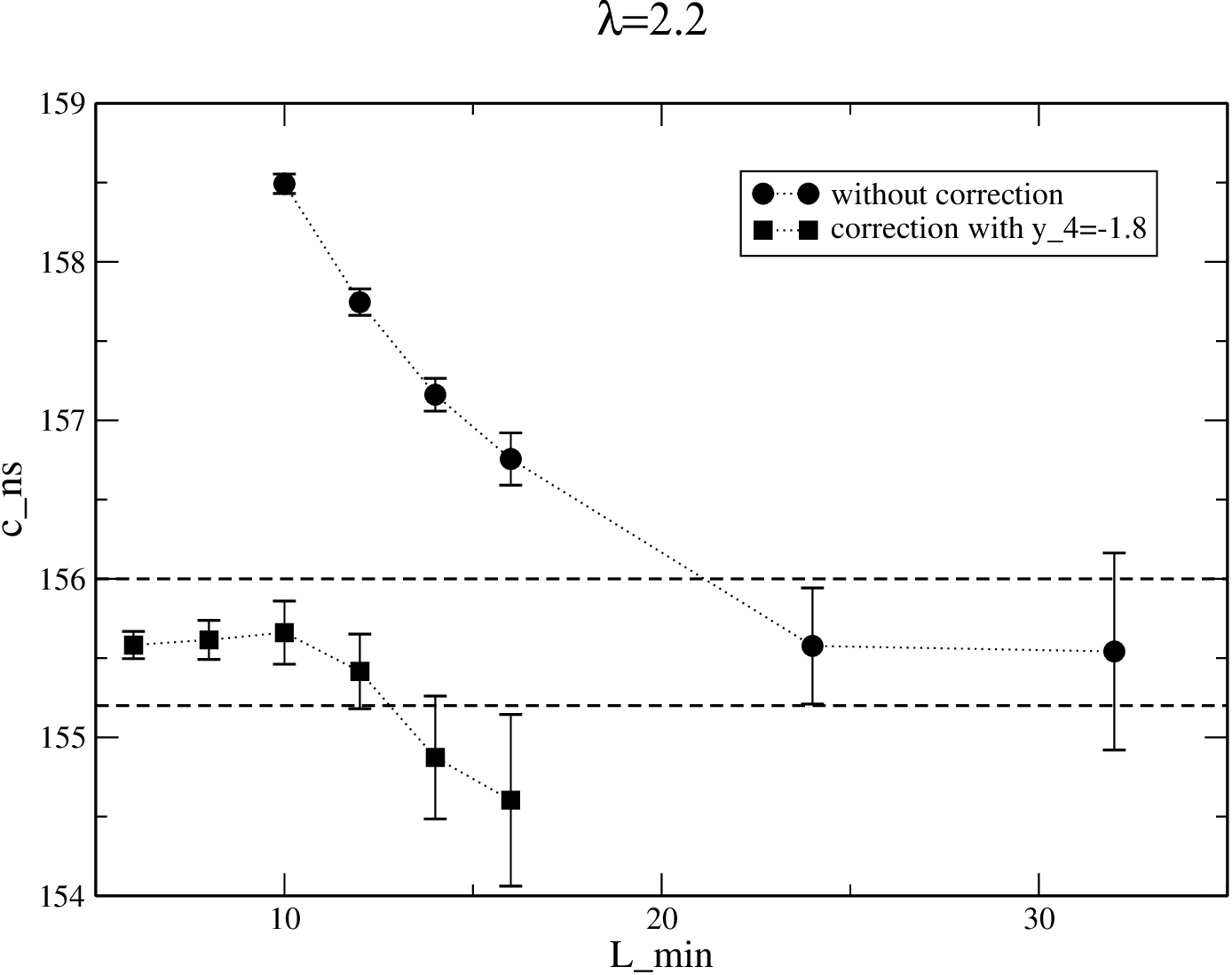}
}
\end{center}
\vskip0.0cm
\caption{\label{cnsplot}
The non-singular part of the specific heat $c_{ns}$ obtained from numerical 
data at $\beta_c$ as a 
function of the smallest lattice size $L_{min}$ that has been included in the 
fit. We have used either an ansatz without any correction to 
scaling~(\ref{cansatzn}) or
an ansatz with subleading corrections to scaling~(\ref{cansatz2s}) with an 
RG-exponent $y_4=-1.8$. For details see the text.
}
\end{figure}

Next we have fitted the specific heat at $\beta_c$ using the ans\"atze
\begin{equation}
\label{cansatzn}
C = c_{ns} + c_s L^{-d+2/\nu}
\end{equation}
and
\begin{equation}
\label{cansatz2s}
C = c_{ns} + c_s L^{-d+2/\nu}  (1+f_2 L^{-\omega_2})
\end{equation}
with $\nu=0.6717$ fixed.
Fitting with ansatz~(\ref{cansatzn}) 
we get $\chi^2$/d.o.f. $\approx 1$ only for $L_{min}>16$   for both values
of $\lambda$. Hence we performed fits with ansatz~(\ref{cansatz2s}) in addition.
We have fixed either $\omega_2=1.8$ or $\omega_2=2.0$. For this type of fit
we get $\chi^2$/d.o.f. $\approx 1$ starting from $L_{min}=6$.  The results
of $c_{ns}$ obtained with $\omega_2=1.8$ and $\omega_2=2.0$ differ only 
little. 
In fig. \ref{cnsplot} we have plotted the results for $c_{ns}$ obtained 
from the fits discussed above. 

Guided by fig. \ref{cnsplot},
we take, for both values of $\lambda$, our final estimate of $c_{ns}$ from the 
fit with ansatz~(\ref{cansatzn})  and $L_{min}=24$. The error, which is 
indicated by dashed lines in fig. \ref{cnsplot}, is chosen such that it covers
the estimates from fits with ansatz~(\ref{cansatzn}) as well as with 
ansatz~(\ref{cansatz2s}).
The dependence of the 
final result on the input values of $\beta_c$ and $\nu$ is estimated by 
redoing the fit with ansatz~(\ref{cansatzn}) and $L_{min}=24$ for 
slightly different values of $\nu$ and $\beta_c$ than those chosen above. 
As result we obtain 
\begin{equation}
\label{c0l2.1}
 c_{ns} = 157.9(5) +  147000 \times (\beta-0.5091503)
                  -2.1 \times (1/\alpha+1/0.0151)
\end{equation}
for $\lambda=2.1$ and
\begin{equation}
\label{c0l2.2}
  c_{ns} = 155.6(4)  + 121000 \times (\beta-0.5083355) 
                  -2.1 \times (1/\alpha+1/0.0151)
\end{equation}
for $\lambda=2.2$. 

\subsection{The energy density in the thermodynamic limit}
The simulations in the high temperature phase were already 
discussed in ref. \cite{we2006}. In the high temperature phase we expect 
that the observables converge exponentially fast to the thermodynamic limit as
$L \rightarrow \infty$.  Throughout we have used $L > 10 \xi$  in these
simulations. In particular for our largest correlation length $\xi \approx 30$
we used $L=350$. 
Hence within our numerical accuracy, the results should coincide with the 
thermodynamic limit.

In the low temperature phase, the breaking of the continuous $U(1)$ symmetry
leads to a Goldstone mode. As a result, leading corrections to 
the thermodynamic limit of the energy density are $O(L^{-3})$ 
\cite{HaLe1990,DiHaNaNi1991}.
We fitted our data for the energy density with the ansatz
\begin{equation}
\label{goldstone}
E(L) = E(\infty) + c L^{-3} \;\;.
\end{equation}
Typically, the difference $E(\infty) - E(L_{max})$ of the fit result for the
thermodynamic limit and the result for the largest lattice size $L_{max}$
that we have simulated is of a similar size as the statistical error of 
$E(L_{max})$ and $E(\infty)$.  
\begin{figure}[tp]
\begin{center}
\scalebox{0.52}
{
\includegraphics{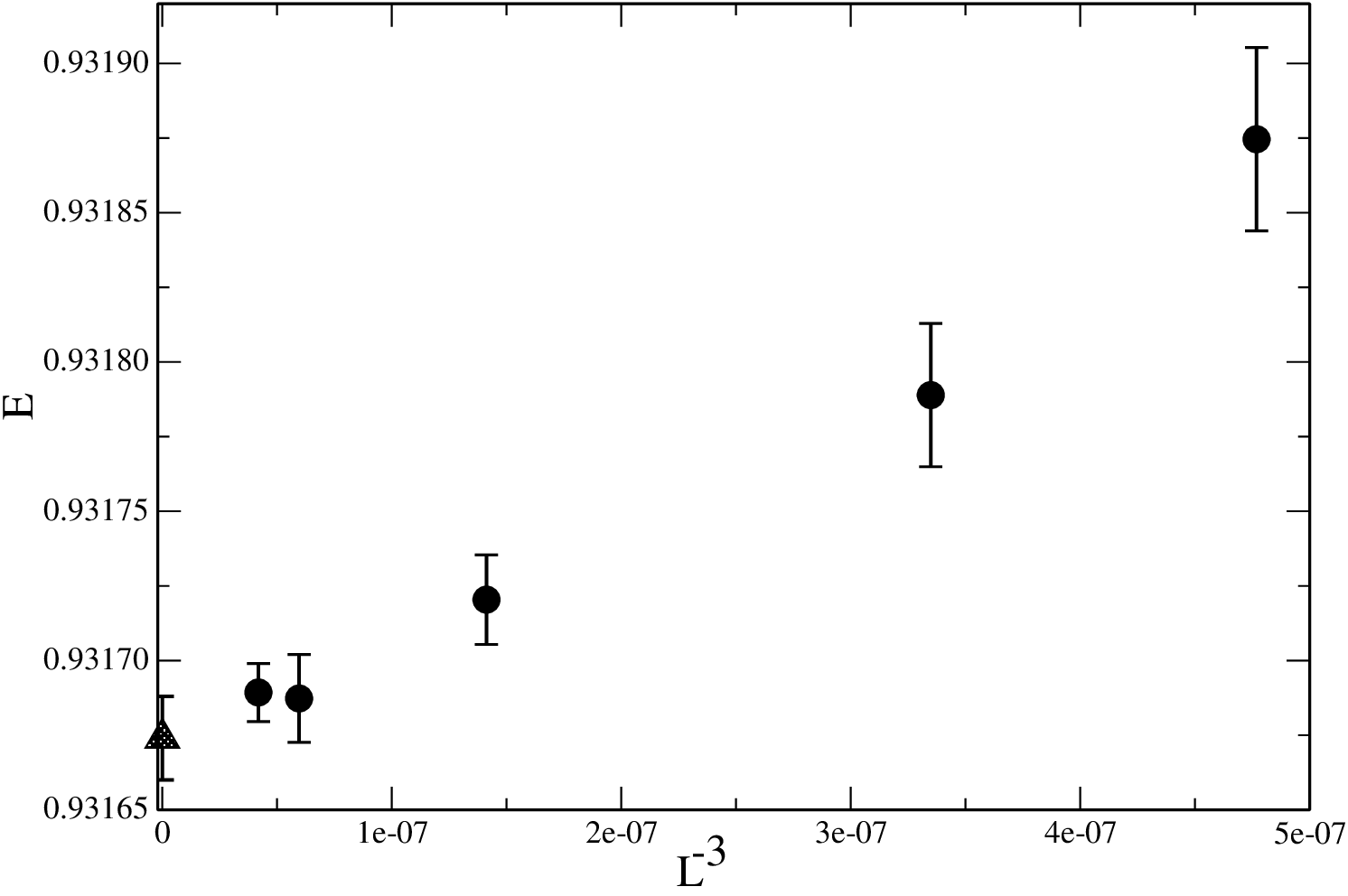}
} \\
\end{center}
\vskip1.0cm
\caption{\label{eextrapol} The energy density $E$ at $\lambda=2.1$ and 
$\beta=0.51$ is plotted as a function of $L^{-3}$, where $L$ is the linear
lattice size. The circles give our Monte Carlo results for the lattice 
sizes $L=128,144,192,256$ and 288.  The triangle gives the result
of the extrapolation to $L=\infty$ with ansatz~(\ref{goldstone}). In the fit
$L=128$ had been excluded.
}
\end{figure}
Therefore we are confident that the quoted error, which 
is the statistical error of the result of the fit with 
ansatz~(\ref{goldstone}), is reliable. For illustration we have plotted in 
Fig.~(\ref{eextrapol})
the Monte Carlo results for the energy density $E$ at 
$\lambda=2.1$ and $\beta=0.51$ as a function of $L^{-3}$. 
Our final results for the thermodynamic limit of the 
energy density are summarized in table \ref{energyoff}. In addition
we give the second moment correlation length $\xi_{2nd}$ in the 
high temperature phase.

We have also measured the specific heat. However the result for $R_{\alpha}$
obtained from these data is consistent with but 
less precise than that obtained from the energy density. Therefore we skip 
the discussion of the specific heat data. 

\begin{table}
\caption{\sl \label{energyoff} Results for the energy density $E$ of the 
$\phi^4$ model at $\lambda=2.1$ and $\lambda=2.2$ in the thermodynamic 
limit for various values of the inverse temperature $\beta$. 
In addition we give second moment correlation length $\xi_{2nd}$ 
in the high temperature phase.
}
\begin{center}
\begin{tabular}{|l|l|l|l|}
\hline
$\lambda$ &\phantom{x} $\beta$ &\phantom{x} $\xi_{2nd}$ &\phantom{xxx} $E$ \\
\hline
2.1 & 0.503  &\phantom{0}8.042(1) & 0.856373(7) \\ 
2.1 & 0.505  &10.482(2) & 0.871351(8) \\
2.1 & 0.506  &12.626(2) & 0.879580(6) \\
2.1 & 0.507  &16.318(4) & 0.888476(7) \\
2.1 & 0.5075 &19.498(6) & 0.893283(9) \\
2.1 & 0.508  &24.845(8) & 0.898418(6) \\
2.1 & 0.5083 &30.453(10)& 0.901727(4) \\
\hline
2.1& 0.51  &        & 0.931674(14) \\
2.1& 0.5105 &      & 0.941232(16) \\ 
2.1& 0.511 &       & 0.950382(13) \\  
2.1& 0.512 &       & 0.967852(11) \\  
2.1& 0.513 &       & 0.984474(10) \\
2.1& 0.515 &       & 1.016005(30) \\
\hline
\hline
2.2& 0.5010  &\phantom{0}7.1723(5)  & 0.849150(4) \\
2.2& 0.5035  &\phantom{0}9.502(1) & 0.866864(5) \\
2.2& 0.5055  &13.610(2) & 0.882972(5) \\ 
2.2& 0.5067  &19.710(5) & 0.894014(6) \\
2.2& 0.50748 &30.475(10)& 0.902171(4) \\
\hline
2.2&  0.5095 &   & 0.937817(19)  \\
2.2&  0.51   &   & 0.947000(19) \\
2.2&  0.511  &   & 0.964370(17) \\
2.2&  0.512  &   & 0.980920(19)  \\
\hline
\end{tabular}
\end{center}
\end{table}

First we have fitted our data for the energy density with the ansatz
\begin{equation}
\label{simpleenergy}
E = e_{ns} + c_{ns} (\beta-\beta_c) + a_{\pm} |\beta-\beta_c|^{1-\alpha} \;\;.
\end{equation}
In these fits, we take $\alpha$, $\beta_c$ and  the corresponding 
values of $e_{ns}$, eqs.~(\ref{e0l2.1},\ref{e0l2.2}), and $c_{ns}$,
eqs.~(\ref{c0l2.1},\ref{c0l2.2}), as input. 

\begin{figure}[tp]
\begin{center}
\scalebox{0.52}
{
\includegraphics{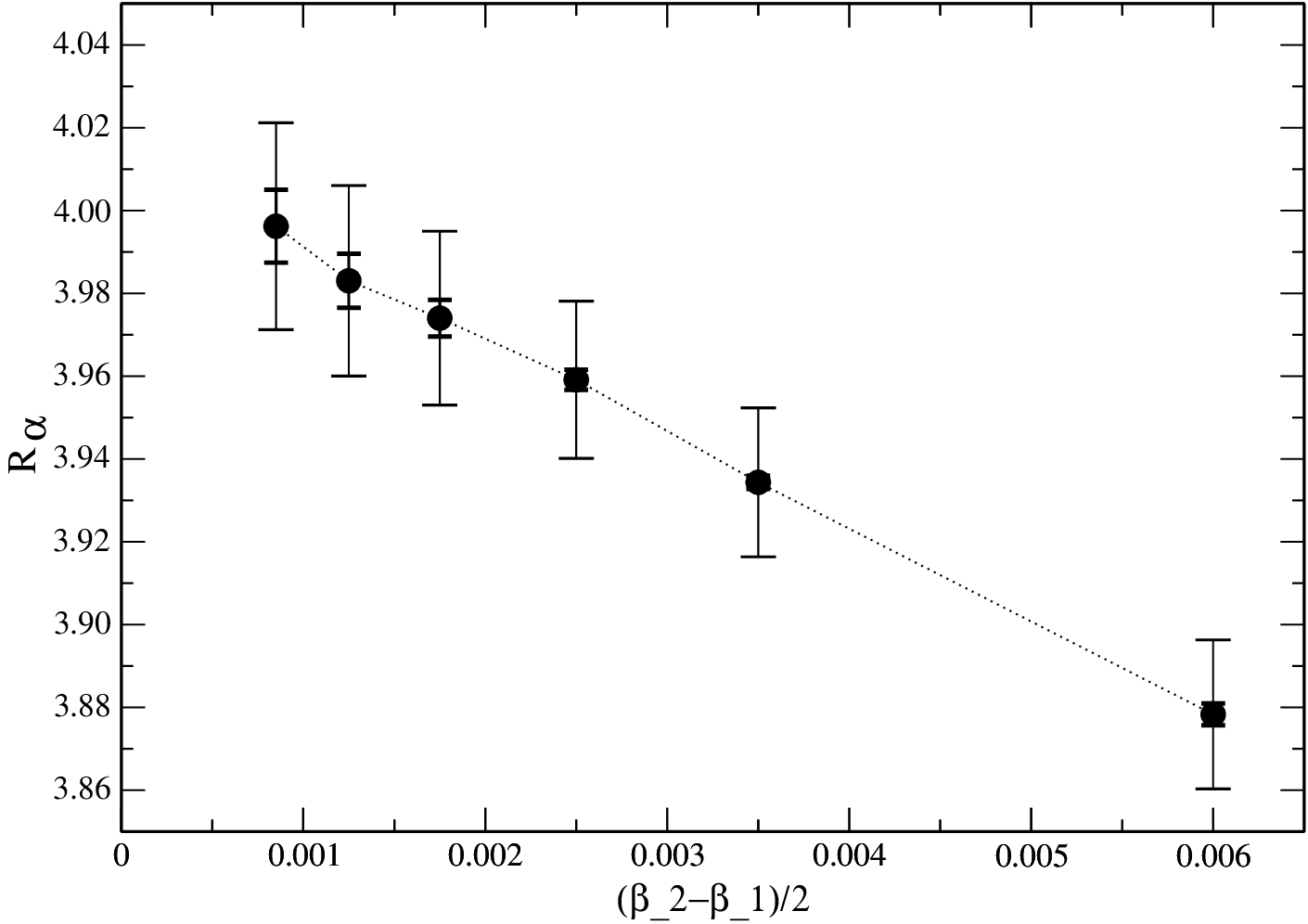}
} \\ 
\end{center}
\vskip1.0cm
\caption{\label{ralphaplot}
Estimate of $R_{\alpha}$ obtained from two values of the inverse temperature
$\beta_1$ and $\beta_2$ using the ansatz~(\ref{simpleenergy}). Only results
for $\lambda=2.1$ are given.
Two errors are displayed: The smaller one is the due to the statistical 
error of the energy density at $\beta_1$ and $\beta_2$; the larger one is due 
to the uncertainty of $c_{ns}$, $e_{ns}$ and $\beta_c$.  The dotted line 
should only guide the eye.
}
\end{figure}

It turns out that in the case of $\lambda=2.1$, fits that include  
$|\beta-\beta_c| \le 0.005$  have a large $\chi^2$/d.o.f. .
To understand this problem we resorted to a simpler analysis of the data. 
We took pairs of inverse temperatures such that 
$\beta_c - \beta_1 \approx \beta_2 - \beta_c$. These two values are sufficient 
to determine the parameters $a_+$  and $a_-$ of the 
ansatz~(\ref{simpleenergy}).  
In Fig.~(\ref{ralphaplot})
we give the results for $R_{\alpha}$ obtained this way as a function of 
$(\beta_2-\beta_1)/2$. 
We see that the estimate of $R_{\alpha}$ is roughly linear in 
$\beta_2-\beta_1$.  A linear extrapolation to $\beta_2-\beta_1=0$ suggests 
a value of $R_{\alpha}$ that is slightly larger than $4$. 

Motivated by this observation, we performed fits with the ansatz
\begin{equation}
\label{simpleenergy2}
E = e_{ns} + c_{ns} (\beta-\beta_c) + a_{\pm} |\beta-\beta_c|^{1-\alpha}
           + d (\beta-\beta_c)^2 \;\;,
\end{equation}
where we have added a further term to the Taylor expansion of the analytic 
part of the energy density.
Note that the first analytic correction to the singular part
comes with a very similar exponent: $2-\alpha$. Hence the fit parameter $d$ 
will provide only an effective amplitude for the combination of the two
terms. Note that also the exponent of subleading corrections 
$1-\alpha+\Delta_2$ is only slightly larger than $2$.
We also tried to explicitely take into account these terms in the fit. 
This leads however to very large errors for the coefficients.

Using ansatz~(\ref{simpleenergy2}) we get fits with a $\chi^2/$d.o.f. smaller 
than one for the interval $0.506 \le \beta \le 0.512$ and 
$0.5055 \le \beta \le  0.511$ for $\lambda=2.1$ and $\lambda=2.2$, respectively.
Using the central values of $c_{ns}$, $e_{ns}$, $\beta_c$ and $\alpha$ 
as input we get 
$R_{\alpha}=4.011(5)$ẹ and $R_{\alpha}=4.017(6)$ for $\lambda=2.1$ and 
$\lambda=2.2$, respectively. Remember that for $\lambda^*=2.15(5)$ leading 
corrections to scaling vanish \cite{we2006}.   Since the results for 
$R_{\alpha}$ at $\lambda=2.1$ and $\lambda=2.2$ are almost the same, we 
conclude that leading corrections to scaling can be safely ignored.

In order to check for the effect of corrections to the 
ansatz~(\ref{simpleenergy2}) discussed above, 
we have repeated the fits using $\beta$-values that are further 
off from $\beta_c$. In particular, fitting the data for 
$\beta=0.503,0.505,0.506,0.512,0.513,0.515$ in the case of $\lambda=2.1$ 
we obtain 
$R_{\alpha}=4.006(3)$ and fitting the data for $\beta=0.501,0.5035,0.5055,
0.511$, $0.512$ in the case of $\lambda=2.2$ we obtain $R_{\alpha}=3.988(2)$.
Note that the average of $|\beta-\beta_c|$ for this second set of fits is 
more than twice as large as for the first set of fits. 

Next we have computed the error of $R_{\alpha}$ due to the uncertainty 
of the input values of $c_{ns},e_{ns}$ and $\beta_c$. To this end, 
we have repeated the fits using the central values of $c_{ns}$, $e_{ns}$
and $\beta_c$ shifted by the error estimate.
Summing the errors of $R_{\alpha}$ due  these input values
we get a little less than $0.02$ for both values of $\lambda$.

Finally, we have also repeated the fits for $\alpha \ne 0.0151$ to  
obtain the dependence of our numerical estimate of $R_{\alpha}$  on 
$\alpha$. 

Summing up all errors discussed above, we arrive at the final estimate 
\begin{equation}
 R_{\alpha}=4.01(5) -  8 \times (\alpha+0.0151) \;\;.
\end{equation}
Notice that the dependence on $\alpha$ is rather small; e.g. inserting 
the estimate $\alpha=-0.0127$ of ref.~\cite{lipa2003} we get 
$R_{\alpha}= 3.99$.

\section{Conclusions and Comparison with the literature}
We have studied the universal amplitude ratio $A_+/A_-$ 
of the specific heat in the three dimensional XY universality class. 
Since $|\alpha|$ is rather small, it is difficult to disentangle the
singular and the non-singular part of the specific heat or the energy
density. This problem holds for numerical data obtained from Monte
Carlo simulations of lattice models as well as for experimental data.

While in ref.~\cite{CEHMS-02} the standard XY model was simulated we
have studied the improved $\phi^4$ model, allowing us to ignore
leading corrections to scaling in the analysis of the data. Given 
the problem discussed above, this is an important advance. 

In our analysis we have combined information from the finite size scaling 
behaviour at the transition \cite{we2006,ourXY}
with precise results for the thermodynamic limit in the low and 
the high temperature phase. Reaching correlation lengths  up to 
$\xi_{2nd} \approx 30$
subleading corrections to scaling are under control.

We made an effort to take carefully into account various
sources of systematic error. Pessimistically we have summed up  these 
errors to arrive at our final estimate 
$R_{\alpha}=(1-A_+/A_-)/\alpha=4.01(5)$.  

Our estimate for $R_{\alpha}$ is significantly smaller than most theoretical
and experimental results. There is only good agreement with the 
experimental result $R_{\alpha}=3.98(2)$ \cite{lich2}.

\section{Acknowledgement}
I like to thank Ettore Vicari for discussions and encouragement.

\end{document}